\documentstyle[preprint,aps]{revtex}

\begin{document}

%\draft

\title{`Baryonic' bound-state instability in trapped fermionic atoms}

\author{A.~S.~Stepanenko and J.~M.~F.~Gunn}

\address{School of Physics and Astronomy, University of Birmingham,
Edgbaston, Birmingham B15 2TT, United Kingdom.}

\date{\today}

\maketitle

\begin{abstract}
We consider a homogeneous gas of spin-$S$ fermionic atoms, as might occur near 
the center of an optical trap. In the case where all scattering lengths are 
negative and of the same magnitude we demonstrate the instability of the Fermi 
sea to the condensation of bound `baryonic' composites containing $2S+1$ 
atoms. The gap in the excitation spectrum is calculated.
\end{abstract}

%\pacs{??}

\section{Introduction}

Sympathetic cooling of trapped
fermionic gases\cite{geist} is likely to lead to the creation of degenerate
Fermi gases in the near future. In magnetic traps, the fermion's spin
is locked to the field direction. However the advent\cite{optical,Cata} of 
optical
traps holds the promise of
degenerate Fermi gases with unquenched spins and hence a variety of nonideal 
Fermi
gases. Already there has been some discussion of the possibility of
pairing\cite{Stoof,Hou1,bar} in the presence of attractive interactions,
which may be of an 
exotic type\cite{Ho2,Ho3}. In part the exotic possibilities occur
because alkali fermions exist which have
larger total spins than spin $1/2$: $^{22}$Na, $^{86}$Rb, $^{132}$Cs,
$^{134}$Cs, $^{136}$Cs have 
$S=5/2$, $5/2$, $3/2$, $7/2$, $9/2$  respectively\cite{nuc}. In this
paper we will point out that in addition to pairing there are other
possible ground states, in particular when the atoms experience spin-independent
attractive interactions. 

One-dimensional models of interacting fermions with spins greater than
one half have been considered for a substantial
time, both for repulsive\cite{Sutherland} and
attractive\cite{Tak,schlot1,schlot2} interactions. In the attractive
case, the ground state has been
found to contain bound states of fermions with up to $2S+1$
constituents (with equality in the absence of an applied `magnetic'
field which would distinguish the different spin states). One may
rationalise these results by noting that the Pauli principle does not
militate against binding extra fermions until a spin state must be
doubly occupied, and hence extra nodes in the spatial wave function
occur. 
In the light of 
these results we will examine whether
the ground state of the {\it three}-dimensional homogeneous $S>1/2$ weakly
interacting attractive Fermi gas is characterised by a condensation of composites
of $2S+1$ fermions. The possible condensation of alpha particles as
four-particle composites in
nuclear matter and at the surface of nuclei has a long history whose
relation to the current work we will discuss at the end of this paper.

\section{Baryonic ground state energy}

In three dimensions we use a variational approach similar to that
used in the original BCS paper\cite{BCS}. We will assume that the
composites (`baryons') have zero centre of mass momentum (as in BCS) and are also
total spin singlets, as in the one-dimensional results. (The latter
can be understood physically as providing the lowest kinetic energy
associated with relative motion in the bound state.) The condensed state, with
all baryons having centre of mass momentum zero, is of
the form (here $n=2S+1$):
$$
|{\rm B}\rangle = \left[ \sum_{\{k_i\}} \varphi
(k_1,\cdots,k_n)c^\dagger_{k_1,\sigma_1} \cdots
c^\dagger_{k_n,\sigma_n}\right]^N
$$
where all the spin states in the baryon, $\sigma_i$, (with
$i=1,\cdots,n$) are different. $\varphi
(k_1,\cdots,k_n)$ is the Fourier transform of the completely symmetric
relative wavefunction $\varphi(r_1,\cdots,r_n)$. Unlike the BCS case,
there are several
grand canonical states, $|{\rm BCS}\rangle$, corresponding to $|{\rm B}\rangle$. 
We will construct the simplest one. 

This family of variational states allows us to use $1/n$ as a small
parameter. (Indeed if the coupling between the atoms were strong, there
would be similarities with work\cite{Witten} on the structure of
baryons in QCD using a $1/N$ expansion.)

Consider the following (non-normalized) ground state:
$$
|\tilde\psi_{\rm b}\rangle = 
\exp\left[
\sum_{k_1,\dots,k_n}
\varphi(k_1,\dots,k_n)c^\dagger_1(k_1)\dots c^\dagger_n(k_n)
\right] |0\rangle
$$

Below we show that the corresponding normalized state has the form
\begin{eqnarray}
|\psi_{\rm b}\rangle &=& 
{\rm e}^{(n-2)N_{\rm b}/2}\prod_{k}u^{n-1}(k)\cdot
\exp\left[
\sum_{k_1,\dots,k_n}
\varphi_\alpha(k_1,\dots,k_n)c^\dagger_1(k_1)\dots c^\dagger_n(k_n)
\right] |0\rangle
\nonumber\\
&=& 
{\rm e}^{(n-2)N_{\rm b}/2}
\prod_{k_1+\dots+k_n=0}
\left[
\prod_{i=1\atop i\neq\alpha}^n u(k_i) 
 + v(k_1,\dots,k_n)c^\dagger_1(k_1)\dots c^\dagger_n(k_n)
\right] |0\rangle
\label{eq1}
\end{eqnarray}
where $c^\dagger_\alpha(k),c_\alpha(k),\alpha=1,\dots,n=2S+1$ the creation and 
annihilation operators of a fermion with $\alpha$-th projection of the spin and 
momentum $k$, $N_{\rm b}\equiv N/n$ is the number of baryons (with $N$ being the 
number of atoms) and
\begin{equation}
\varphi_\alpha(k_1,\dots,k_n)
\equiv \frac{v(k_1,\dots,k_n)u(k_\alpha)}{u(k_1)\cdot\dots\cdot u(k_n)}
\propto\delta_{k_1+\dots+k_n,0}
\label{eq2}
\end{equation}
The Kronecker symbol on the r.h.s. of (\ref{eq2}) means that we consider 
baryonic states with total momentum equal to zero. The expression 
$k_1+\dots+k_n=0$ as a subscript to the product in (\ref{eq1}) implies that the 
product is taken over all sets $\{k_1,\dots,k_n\}$ with the total momentum equal 
to zero. We will apply a `normalization' condition similar to that used in BCS 
states:
\begin{equation}
|u(k)|^2 + |v(k)|^2 = 1\ ,\quad
|v(k)|^2\equiv\sum_{k_2,\dots,k_n}|v(k,k_2,\dots,k_n)|^2
\label{eq3}
\end{equation}
This, unlike the BCS one, does not imply that the total state is normalised.
The vector conjugate to $|\psi_{\rm b}\rangle$  is
\begin{equation}
\langle\psi_{\rm b}|
 = 
{\rm e}^{(n-2)N_{\rm b}/2}\prod_{k}\bar u^{n-1}(k)\cdot
\langle 0|\exp\left[
\sum_{k_1,\dots,k_n}
\bar\varphi_\alpha(k_1,\dots,k_n)c_n(k_n)\dots c_1(k_1)
\right] 
\label{eq4}
\end{equation}
To calculate the normalization $\langle\psi_{\rm b}|\psi_{\rm b}\rangle$ we make 
use of the following identity
\begin{eqnarray}
&&\exp\left[
\sum_{k_1,\dots,k_n}\varphi_\alpha(k_1,\dots,k_n)
c^\dagger_1(k_1)\dots c^\dagger_n(k_n)
\right]
\nonumber\\
&&\qquad
 = \int {\rm D}\bar\xi{\rm D}\xi\ \exp\Biggl[
\sum_{i=1}^{n-1}\sum_{k}\left\{ - \bar\xi_i(k)\xi_i(k)
 + c^\dagger_i(k)\xi_i(k)\right\}
\nonumber\\
&&\qquad\quad
 - \sum_{k_1,\dots,k_n}\varphi_\alpha(k_1,\dots,k_n) c^\dagger_n(k_n) 
\bar\xi_1(k_1)\dots \bar\xi_{n-1}(k_{n-1})
\Biggr]
\label{eq5}
\end{eqnarray}
where $\bar\xi_i(k),\xi_i(k)$ are Grassmann variables and the measure of 
integration is denoted by
$$
{\rm D}\bar\xi{\rm D}\xi\equiv
\prod_{i=1}^{n-1}\prod_{k}{\rm d}\bar\xi_i(k){\rm d}\xi_i(k)
$$
As $c^\dagger$ enters linearly in the exponentials in (\ref{eq5}) then the 
operator averaging in $\langle\psi_{\rm b}|\psi_{\rm b}\rangle$ can be easily 
fulfilled and we find
\begin{eqnarray*}
\langle\psi_{\rm b}|\psi_{\rm b}\rangle &=& 
C \int {\rm D}\bar\xi{\rm D}\xi\ \exp\Biggl[
 - \sum_{i=1}^{n-1}\sum_{k}\bar\xi_i(k)\xi_i(k)
\\
&-&
   \sum_{k_1,\dots,k_n\atop p_1,\dots,p_n}
   \bar\varphi_\alpha(k_1,\dots,k_n)\varphi_\alpha(p_1,\dots,p_n) 
   \bar\xi_1(p_1)\xi_1(k_1)\dots \bar\xi_{n-1}(p_{n-1})\xi_{n-1}(k_{n-1})
   \delta_{k_n,p_n}
\Biggr]
\end{eqnarray*}
where $C$ denotes the factor
$$
C = {\rm e}^{(n-2)N_{\rm b}}\prod_{k}|u(k)|^{n-1}
$$
For the second term in the exponential we make use of the following 
identity:
\begin{eqnarray}
&&\!\!\!
\exp\Biggl[
 - \sum_{k_1,\dots,k_n\atop p_1,\dots,p_n}
   \bar\varphi_{n}(k_1,\dots,k_n)\varphi_{n}(p_1,\dots,p_n) 
   \bar\xi_1(p_1)\xi_1(k_1)\dots \bar\xi_{n-1}(p_{n-1})\xi_{n-1}(k_{n-1})
   \delta_{k_n,p_n}
\Biggr]
\nonumber\\
&& = 
\int{\rm D}\bar\sigma{\rm D}\sigma\
\exp\Biggl[
 - \sum_{i=1}^{n-1}\sum_{k,p}\bar\sigma_i(k,p)\sigma_i(k,p)
 - \sum_{i=1}^{n-1}\sum_{k,p}\bar\sigma_i(k,p)\bar\xi_i(k)\xi_i(p)
\nonumber\\
&&\qquad
 + \sum_{k_1,\dots,k_n\atop p_1,\dots,p_n}
   \bar\varphi_{n}(k_1,\dots,k_n)\varphi_{n}(p_1,\dots,p_n) 
   \prod_{i=1}^{n-1}\sigma_i(p_i,k_i) \delta_{k_n,p_n}
\Biggr]
\label{sigma}
\end{eqnarray}
where
$$
{\rm D}\bar\sigma{\rm D}\sigma\equiv
\prod_{i=1}^{n-1}\prod_{k,p}{\rm d}\bar\sigma_i(k,p){\rm d}\sigma_i(k,p)
$$
The meaning of $\sigma$-fields is not clear at this point, but we will discuss 
this presently.

Integrating over the Grassmann fields we obtain
$$
\langle\psi_{\rm b}|\psi_{\rm b}\rangle = 
C \int{\rm D}\bar\sigma{\rm D}\sigma\
\exp\Biggl[
 - \sum_{i=1}^{n-1}\sum_{k,p}\bar\sigma_i(k,p)\sigma_i(k,p)
 + \sum_{i=1}^{n-1}{\rm tr}\log[1+\bar\sigma_i]
 + K
\Biggr]
$$
where $K$ denotes the following expression
$$
K = 
   \sum_{k_1,\dots,k_n\atop p_1,\dots,p_n}
   \bar\varphi_{n}(k_1,\dots,k_n)\varphi_{n}(p_1,\dots,p_n) 
   \prod_{i=1}^{n-1}\sigma_i(p_i,k_i) \delta_{k_n,p_n}
$$
We calculate the integral over $\bar\sigma,\sigma$ using the saddle point method, 
where $n$ is the large parameter. 
We assume that the symmetry of the exponential over $\sigma_i$-fields is not 
broken and put $\sigma_i=\sigma,\bar\sigma_i=\bar\sigma$, so the exponent becomes
$$
 - (n-1)\bar\sigma\sigma
 + (n-1){\rm tr}\log[1+\bar\sigma]
 + K
$$
The saddle point equations are
\begin{eqnarray}
\sigma(k,p) &=& (1+\bar\sigma)^{-1}(k,p)
\label{sp1}\\
\bar\sigma(k,p)
 &=& 
   \sum_{k_2,\dots,k_n\atop p_2,\dots,p_n}
   \bar\varphi_{n}(k,k_2,\dots,k_n)\varphi_{n}(p,p_2,\dots,p_n) 
   \prod_{i=2}^{n-1}\sigma(p_i,k_i) \delta_{k_n,p_n}
\label{sp2}
\end{eqnarray}
In the limit $n\to\infty$ we expect a mean field approximation to be 
valid for the description of the composites:
\begin{equation}
v(k_1,\dots,k_n) = B\cdot\delta_{k_1+\dots+k_n,0}
\prod_{i=1}^n v(k_i)
\label{eq16}
\end{equation}
where $B$ is chosen to satisfy the normalization condition (\ref{eq3}). With this 
assumed form for $v$, it can be shown that the following anzatz can serve as a 
quite general solution of the saddle point equations:
\begin{equation}
\sigma(k,p) = \delta_{k,p}s_1(k) + c_2\bar s_2(k)s_2(p)\ ,\qquad
\bar\sigma(k,p) = \delta_{k,p}s_3(k) + c_4\bar s_4(k)s_4(p)\ .
\label{anz}
\end{equation}
with functions $s_i,i=1,2,3,4$ to be determined.
If we further restrict ourselves by considering weak coupling case we obtain that 
the solution is (see Appendix for details)
\begin{equation}
\sigma(k,p) = \delta_{k,p}|u(k)|^2\ ,\qquad
\bar\sigma(k,p) = \delta_{k,p}\left|\frac{v(k)}{u(k)}\right|^2
\label{sol}
\end{equation}
Hence,
$$
\langle\psi_{\rm b}|\psi_{\rm b}\rangle = 
C \exp\Biggl[
 - (n-1)\sum_{k}|v(k)|^2
 - (n-1)\sum_{k}\log|u(k)|^2
 + \sum_k\sum_{k_2,\dots,k_n} |v(k,k_2,\dots,k_n)|^2
\Biggr] 
$$
Because of the normalization condition (\ref{eq3}) and expression for $C$ we 
finally obtain
\begin{equation}
\langle\psi_{\rm b}|\psi_{\rm b}\rangle = 1 + {\rm O}(1/n)
\label{eq9}
\end{equation}

Let us consider the following average:
$$
\langle\psi_{\rm b}|c_\alpha(k)c^\dagger_\alpha(k')|\psi_{\rm b}\rangle
$$
Repeating all the steps we have made before we can obtain the following result:
\begin{eqnarray*}
\langle\psi_{\rm b}|c_\alpha(k)c^\dagger_\alpha(k')|\psi_{\rm b}\rangle
 &=& C \int{\rm D}\bar\sigma{\rm D}\sigma\ (1+\bar\sigma_\alpha)^{-1}(k,k')
\\
 && \times
\exp\Biggl[
 - \sum_{i=1}^{n-1}\sum_{k,p}\bar\sigma_i(k,p)\sigma_i(k,p)
 + \sum_{i=1}^{n-1}{\rm tr}\log[1+\bar\sigma_i]
 + K
\Biggr]
\end{eqnarray*}
Calculating again the integral over $\sigma$-fields using the saddle point method 
and noting that the contribution of the first multiplier in the integrand to the 
saddle point equations can be neglected we obtain from (\ref{sp1}):
$$
\langle\psi_{\rm b}|c_\alpha(k)c^\dagger_\alpha(k')|\psi_{\rm b}\rangle
 = \sigma(k,k')
$$
So the $\sigma$-field may be interpreted as momentum distribution of the holes.

In an analogous way we can obtain the following averages:
\begin{equation}
\langle\psi_{\rm b}|c^\dagger_\alpha(k)c_\alpha(k)|\psi_{\rm b}\rangle
 = |v(k)|^2 + {\rm O}(1/n)
\label{eq10}
\end{equation}
\begin{eqnarray}
&&\!\!\!\!
\sum_q\langle\psi_{\rm b}|c^\dagger_\alpha(k)c^\dagger_\beta(-k+q)
c_\beta(-k'+q)c_\alpha(k')|\psi_{\rm b}\rangle
\nonumber\\
&&\qquad
 = \delta_{k,k'} |v(k)|^2\sum_q |v(q)|^2
 + \bar v(k) u(k) \bar u(k') v(k') f(\hat k,\hat k') + {\rm O}(1/n)
\label{eq11}
\end{eqnarray}
where $\hat k\equiv\frac{k}{|k|}$ and the function $f$ has the form
\begin{equation}
f(\hat k,\hat k')\equiv f(x) = 1 - \frac{3x}{2} + \frac{x^3}{2}\ ,\qquad
x\equiv \sin\frac{\theta}{2}
\label{f}
\end{equation}
with $\theta$ being the angle between vectors $k,k'$.

We are now in a position to consider a general Hamiltonian of the form:
\begin{equation}
\hat H =
\sum_{k,\alpha} \epsilon(k) c^\dagger_\alpha(k) c_\alpha(k)
 + \frac{1}{2\Omega}\sum_{k,k',q\atop\alpha,\beta}
   V(k,k') c^\dagger_\alpha(k)c^\dagger_\beta(-k+q)c_\beta(-k'+q)c_\alpha(k')
\label{eq12}
\end{equation}
where
\begin{equation}
\epsilon(k) = \frac{k^2}{2m} - \mu
\end{equation}
and $\mu$ is the chemical potential. Then, using the results above for
expectation values, we find the following expression for 
the ground state energy
\begin{equation}
E_{\rm b} = \langle\psi_{\rm b}|\hat H|\psi_{\rm b}\rangle = 
   n\sum_{k} \epsilon(k) |v(k)|^2 + \frac{V(0)N^2}{2\Omega}
 + \frac{n^2}{2\Omega}\sum_{k,k'}V(k,k') \bar v(k) u(k) \bar u(k') v(k') 
   f(\hat k,\hat k')
\label{eq13a}
\end{equation}
with the condition
\begin{equation}
\langle\psi_{\rm b}|\hat N|\psi_{\rm b}\rangle = n \sum_{k} |v(k)|^2 = N
\label{eq14}
\end{equation}
Note that if we consider an interaction potential depending only on the modulus 
of momentum: $V(k,k')=V(|k|,|k'|)$, and look for a solution for $u,v$ also 
depending only on the modulus of momentum (that is in absence of a spontaneous 
breaking of the rotational symmetry) then the function $f$ in (\ref{eq13a}) can 
be replaced by its average value:
$$
\int_0^1{\rm d}x\ f(x) = \frac{3}{8}
$$
and we have
\begin{equation}
E_{\rm b} = \langle\psi_{\rm b}|\hat H|\psi_{\rm b}\rangle = 
   n\sum_{k} \epsilon(k) |v(k)|^2 + \frac{V(0)N^2}{2\Omega}
 + \frac{n^2}{2\Omega}\sum_{k,k'}\tilde V(k,k') \bar v(k) u(k) \bar u(k') v(k')
\label{eq13}
\end{equation}
where $\tilde V = \frac{3}{8}V$. 

The normalization condition, (\ref{eq3}), allows the introduction of the 
following parameterization for $u,v$, by analogy with the usual procedure for
the BCS case:
\begin{equation}
v(k) = \cos\theta(k)\ ,\qquad u(k) = \sin\theta(k)
\label{eq15}
\end{equation}
and we obtain for the ground state energy
\begin{equation}
E_{\rm b} = n\sum_{k} \epsilon(k) \cos^2\theta(k) + \frac{V(0)N^2}{2\Omega}
 + \frac{n^2}{2\Omega}\sum_{k,k'}V(k,k') 
   \cos\theta(k)\sin\theta(k)\cos\theta(k')\sin\theta(k')
\label{eq21}
\end{equation}

We now minimize $E_{\rm b}$ with respect to $\theta(k)$ to find
\begin{equation}
\tan[2\theta(k)] = \frac{1}{2\epsilon(k)}\frac{n}{(2\pi)^3}
\int{\rm d}k'\ \tilde V(k,k')\sin[2\theta(k')]
\label{eq23}
\end{equation}
Introducing the following notation
\begin{equation}
\Delta(k) = - \frac{n}{2(2\pi)^3} \int{\rm d}k'\ \tilde V(k,k')
\sin[2\theta(k')]\ ,
\quad E(k) = \left[\epsilon^2(k)+\Delta^2(k)\right]^{1/2}
\label{eq24}
\end{equation}
such that
\begin{equation}
\tan[2\theta(k)] = - \frac{\Delta(k)}{\epsilon(k)}\ ,\quad
\sin[2\theta(k)] = \frac{\Delta(k)}{E(k)}\ ,\quad
\cos[2\theta(k)] = - \frac{\epsilon(k)}{E(k)}
\label{eq25}
\end{equation}
we obtain the following equation for the gap $\Delta(k)$:
\begin{equation}
\Delta(k) = - \frac{n}{2(2\pi)^3}\int{\rm d}k'\ \tilde V(k,k') 
\frac{\Delta(k')}{E(k')}
\label{eq26}
\end{equation}
and the chemical potential can be defined from
\begin{equation}
\frac{n}{(2\pi)^3}\int{\rm d}k'\ \cos^2\theta(k) = \rho
\label{eq26a}
\end{equation}

Following Anderson and Morel\cite{AM}, Eq.(\ref{eq24}) can be rewritten in terms 
of an effective potential, 
$U_\xi(k,k')$:
\begin{equation}
\Delta(k) = - \frac{1}{(2\pi)^3}\int_{|\epsilon(k')|<\xi}{\rm d}k'\ 
U_\xi(k,k')\frac{\Delta(k')}{2E(k')}\ ,
\label{eq28}
\end{equation}
where $\xi$ is some cut-off such that $\xi\ll\mu\equiv\frac{k_{\rm F}^2}{2m}$ 
and $U_\xi(k,k')$ satisfies the equation:
\begin{equation}
U_\xi(k,k') = n\tilde V(k,k')
 - \frac{1}{(2\pi)^3}\int_{|\epsilon(q)|>\xi}{\rm d}q\ 
n\tilde V(k,q)U_\xi(q,k')\frac{1}{2|\epsilon(q)|}\ .
\label{eq29}
\end{equation}

We take $V(k,k')$ to have the following separable (energy dependent) form:
\begin{equation}
V(k,k') = V\Theta(\mu-|\epsilon(k)|)\Theta(\mu-|\epsilon(k')|)\ ,\qquad
V = \frac{4\pi a}{m}
\label{eq30}
\end{equation}
which, as we shall see, is consistent with low-energy approximation to
the $T$-matrix. Here $a$ is 
the scattering length (we put $\hbar=1$), 
$\Theta(x)$ the step function. We assume that the interaction is weak (which is 
a good approximation experimentally) which means 
\begin{equation}
k_{\rm F}a\ll 1
\label{eq31}
\end{equation}
To check that the form of interaction (\ref{eq30}) is a consistent low-energy 
approximation consider the equation for the $T$-matrix $T(k,k',z)$ of zero 
energy, $z=0$ :
\begin{equation}
T(k,k',0) = V(k,k') - \frac{1}{(2\pi)^3}\int_{|\epsilon(q)|>\xi}{\rm d}q\ 
V(k,q)T(q,k',0)\frac{m}{q^2}
\label{eq32}
\end{equation}
Making the anzatz
$$
T(k,k',0) = T_0\Theta(\mu-|\epsilon(k)|)\Theta(\mu-|\epsilon(k')|)
$$
we obtain the following relation for $T_0,V$:
$$
\frac{1}{\rho_0T_0} = \frac{1}{\rho_0V} + \sqrt{2}\ ,\quad
\rho_0 = \frac{mk_{\rm F}}{2\pi^2}
$$
As 
$$
\frac{1}{\rho_0V} = \frac{\pi}{2ak_{\rm F}}\gg1
$$
then
\begin{equation}
T_0 \approx V 
\label{eq33}
\end{equation}

Let us define $U_\xi(k,k')$. Assuming the form
$$
U_\xi(k,k') = U_\xi\Theta(\mu-|\epsilon(k)|)\Theta(\mu-|\epsilon(k')|)
$$
we get the following equation for $U_\xi$:
$$
U_\xi = n\tilde V - \frac{n\tilde VU_\xi}{4\pi^2}
\int\limits_{\xi<|\epsilon(q)|<\mu}{\rm d}q\ \frac{q^2}{|\epsilon(q)|}
$$
After some algebra we arrive at the relation
\begin{equation}
\frac{1}{\rho_0U_\xi} = \frac{1}{\rho_0n\tilde V} + \log(\sqrt{2}+1)
 - \log\frac{\xi}{4\mu}
\label{eq34}
\end{equation}
If we assume again that we can drop the second term in the right hand side 
(because the coupling is weak) and noting that $\tilde V,U_\xi<0$ we finally 
obtain
\begin{equation}
\frac{1}{\rho_0|U_\xi|} = \frac{1}{\rho_0n|\tilde V|} + \log\frac{\xi}{4\mu}
\label{eq35}
\end{equation}

Note that the energy can be expressed in terms of (\ref{eq24}):
\begin{equation}
E_{\rm b} = \frac{n\Omega}{2(2\pi)^3}\int{\rm d}k\ \left[
\epsilon(k) - \frac{\epsilon^2(k)}{E(k)} - \frac{\Delta^2(k)}{2E(k)}
\right]
\label{eq27}
\end{equation}
From (\ref{eq28}) then we obtain
\begin{equation}
\Delta = 2\xi\exp\left[-\frac{1}{\rho_0|U_\xi|}\right]
 = 8\mu\exp\left[-\frac{1}{\rho_0n|\tilde V|}\right]
\label{eq36}
\end{equation}
The energy of the normal state per unit volume can be written as
\begin{equation}
\frac{E_{\rm n}}{\Omega}
 = \frac{n}{(2\pi)^3}\int_{|k|<k_{\rm F}}{\rm d}k\ \epsilon(k)
\label{eq37}
\end{equation}
Hence, from (\ref{eq27},\ref{eq36}) we find that the difference of the baryonic 
and 
the normal ground state energies is:
\begin{equation}
\frac{\Delta E_{\rm b}}{\Omega} \equiv \frac{E_{\rm b} - E_{\rm n}}{\Omega} 
 = - 48\rho\mu\exp\left[-\frac{1}{\rho_0n|\tilde V|}\right]\ ,\quad
\rho\equiv\frac{N}{\Omega}\ ,\quad
\rho_0 = \frac{mk_{\rm F}}{2\pi^2}
\label{eq38}
\end{equation}
with $N$ the number of atoms in the system. 

%We note that we may define an order parameter by analogy with the BCS case, as 
%follows
%\begin{eqnarray}
%\Delta(k_1,\dots,k_n)&\equiv&
% - \frac{n(n-1)}{2(2\pi)^3}\mathop{\rm Sym}_{k_1,\dots,k_n}
%\int{\rm d}k\ V(k_1,k)\langle\psi_{\rm b}|c_1(k)c_2(k_1+k_2-k)\dots 
%c_n(k_n) |\psi_{\rm b}\rangle
%\nonumber\\
%&=&
% - \frac{n(n-1)}{2(2\pi)^3}\mathop{\rm Sym}_{k_1,\dots,k_n}
%\int{\rm d}k\ V(k_1,k)\bar u(k)v(k,k_1+k_2-k,k_3,\dots,k_n) 
%\label{order_par}
%\end{eqnarray}
%where Sym means symmetrization over all momenta $k_1,\dots,k_n$.

We conclude this section with a note about weakening the assumption of 
equality of scattering lengths in all channels. If the scattering lengths are 
different then the interaction will take the form~\cite{Ho1}:
$$
V = \sum_{n=0}^{[f]} V_n ({\bf S}_1\cdot{\bf S}_2)^n
$$
where $[f]$ means the integer part of $f$. In the case where the
scattering 
lengths in 
all channels are equal, $a_F=a,\forall F$, we have $V_0=\frac{4\pi a}{m}, 
V_n=0,n\neq0$. If the scattering lengths are slightly different then
the terms 
in the 
interaction with $V_n,n\neq0$ are small and can be neglected but $V_0$ is equal 
to some weighted average over the scattering lengths and the treatment in 
this section will still be approximately valid.

\section{BCS ground state energy}

We will now show that the conventional BCS ground state has an 
energy that is higher than the baryonic one, at least if we restrict
ourselves to  $s$-wave pairing. The BCS ground state has the 
form~\cite{Voll}:
\begin{equation}
|\psi_{\rm BCS}\rangle = \prod_{k>0}u^n(k)
\exp\left[
\sum_{k>0\atop\alpha,\beta}\varphi(k)c^\dagger_\alpha(k)
\tilde P_{\alpha\beta}c^\dagger_\beta(-k)
\right]|0\rangle
\label{eq39}
\end{equation}
and the conjugate state is
\begin{equation}
\langle\psi_{\rm BCS}| = \prod_{k>0}\bar u^n(k)
\langle0|\exp\left[
 - \sum_{k>0\atop\alpha,\beta}\bar\varphi(k)c_\alpha(k)
\tilde P_{\alpha\beta}c_\beta(-k)
\right]
\label{eq40}
\end{equation}
where
\begin{equation}
\varphi(k)\equiv\frac{v(k)}{u(k)}\ ,\qquad |v(k)|^2 + |u(k)|^2 = 1
\label{eq41}
\end{equation}
$k>0$ is an arbitrary ordering on momentum space which divides it
into two halves 
(it 
can be defined, for example, as follows: $k>0$ if $k_z>0$ or $k_z=0,k_y>0$ or 
$k_z=0,k_y=0,k_x>0$). The matrix $\tilde P$ has the form
\begin{equation}
\tilde P^{\rm T} = - \tilde P
\label{eq42}
\end{equation}
T means matrix transposition. 

It is well known that \cite{Gant} any antisymmetric matrix $A: A^{\rm T}=-A$ can 
be orthogonally transformed to an antisymmetric matrix $B$ in the canonical 
form: 
$$
A = OBO^{\rm T}\ ,\qquad O^{\rm T} = O^{-1}
$$
matrix $B$ has the following quasi-diagonal (block-diagonal) form
$$
B = {\rm diag}\left\{
\begin{array}{cc}
0 & \lambda_1 \\
-\lambda_1 & 0
\end{array}
, \dots ,
\begin{array}{cc}
0 & \lambda_{n/2} \\
-\lambda_{n/2} & 0
\end{array}
\right\}
$$
This transformation is equivalent to
\begin{equation}
c(k)\rightarrow Oc(k)\ ,\qquad c^\dagger(k)\rightarrow c^\dagger(k) O^{\rm T}
\label{eq44}
\end{equation}
As the Hamiltonian in (\ref{eq12}) is invariant under the transformation 
(\ref{eq44}) 
and because of the assumption of the spin symmetry of the ground state, there 
exists some 
orthogonal matrix $O$ such that the ground state (\ref{eq39},\ref{eq40}), after 
transformation (\ref{eq44}), takes the following form:
\begin{equation}
|\psi_{\rm BCS}\rangle = \prod_{k>0}u^n(k)
\exp\left[
\sum_{k>0\atop\alpha,\beta}\varphi(k)c^\dagger_\alpha(k) 
P_{\alpha\beta}c^\dagger_\beta(-k)
\right]|0\rangle
\label{eq45}
\end{equation}
where
\begin{equation}
P = {\rm diag}\left\{
\begin{array}{cc}
 0 & {\rm e}^{{\rm i}\chi_1} \\
-{\rm e}^{{\rm i}\chi_1} & 0
\end{array}
, \dots ,
\begin{array}{cc}
 0 & {\rm e}^{{\rm i}\chi_{n/2}} \\
-{\rm e}^{{\rm i}\chi_{n/2}} & 0
\end{array}
\right\}
\label{eq46}
\end{equation}
The form of the matrix $P$ implies that the spin states  can be 
enumerated after the transformation in such a way that there are $n/2$ 
``positively'' directed spins and each of them has an oppositely directed 
partner. So the spin indices can be thought as taking values $\pm1,\dots,\pm 
n/2$. The phases $\chi_\alpha$ represent possible phase differences between 
Cooper 
pairs in different spin states (we consider only rotationally invariant states, 
so the modulus of all elements in the matrix $P$ are equal).

To calculate the norm and the energy of the ground state we consider the 
following generating functional:
\begin{equation}
G(\bar\eta,\eta) = 
\langle\psi_{\rm BCS}|
\exp\left[\sum_{k,\alpha}c^\dagger_\alpha(k)\eta_\alpha(k)\right]
\exp\left[\sum_{k,\alpha}\bar\eta_\alpha(k)c_\alpha(k)\right]
|\psi_{\rm BCS}\rangle
\label{eq47}
\end{equation}
Making use of the following identity
\begin{eqnarray*}
\exp\left[
\sum_{k>0\atop\alpha,\beta}\varphi(k)c^\dagger_\alpha(k)P_{\alpha\beta} 
c^\dagger_\beta(-k)
\right] &=& \int{\rm D}\bar\xi{\rm D}\xi\ \exp\Biggl[
 - \sum_{k>0,\alpha} \bar\xi_\alpha(k)\xi_\alpha(k)\\
&& + \sum_{k>0,\alpha} c^\dagger_\alpha(k)\xi_\alpha(k)
 + \sum_{k>0\atop\alpha.\beta} \bar\xi_\alpha(k)P_{\alpha\beta}c_\beta(-k)
\Biggr]
\end{eqnarray*}
we get the following result for the generating functional
\begin{eqnarray}
G(\bar\eta,\eta) &=& \exp\Biggl[
\sum_{k}\sum_{\alpha=1}^{n/2} 
\left\{
   \bar u(k)v(k){\rm e}^{{\rm i}\chi_\alpha}
   \bar\eta_\alpha(k)\bar\eta_{-\alpha}(-k)
 - u(k)\bar v(k){\rm e}^{-{\rm i}\chi_\alpha}\eta_\alpha(k)\eta_{-\alpha}(-k)
\right\}
\nonumber\\
&& 
 - \sum_{k}\sum_{\alpha} 
|v(k)|^2\bar\eta_\alpha(k)\eta_{\alpha}(k)
\Biggr]
\label{eq48}
\end{eqnarray}
Putting $\bar\eta,\eta=0$ in (\ref{eq48}) we obtain $\langle\psi_{\rm BCS}| 
\psi_{\rm BCS}\rangle=1$. The energy of the ground state is
\begin{eqnarray}
E_{\rm BCS} &=& 
\langle\psi_{\rm BCS}|\hat H|\psi_{\rm BCS}\rangle 
\nonumber\\
&=& 
   n\sum_{k} \epsilon(k) |v(k)|^2 + \frac{V(0)N^2}{2\Omega}
 + \frac{n}{2\Omega}\sum_{k,k'}V(k,k') 
   u(k)\bar v(k)\bar u(k')v(k')
\label{eq49}
\end{eqnarray}

Introducing as usual the trigonometric parameterization
$$
u(k) = \sin\theta(k)\ ,\qquad v(k) = \cos\theta(k)
$$
and applying the variational principle we find the following equation
\begin{equation}
\tan[2\theta(k)] = \frac{1}{2\epsilon(k)}\frac{1}{(2\pi)^3}
\int{\rm d}k'\ V(k,k')\sin[2\theta(k')]
\label{eq50}
\end{equation}
Defining the gap as
\begin{equation}
\Delta(k) = - \frac{1}{2(2\pi)^3} \int{\rm d}k'\ V(k,k')\sin[2\theta(k')]
\label{eq51}
\end{equation}
and repeating calculations of the preceding section we arrive at the following 
expression for the gap
\begin{equation}
\Delta = 8\mu\exp\left[-\frac{1}{\rho_0|V|}\right]
\label{eq52}
\end{equation}
and the difference between BCS and the normal ground state energies is
\begin{equation}
\frac{\Delta E_{\rm BCS}}{\Omega} \equiv \frac{E_{\rm BCS} - E_{\rm n}}{\Omega} 
 = - 48\rho\mu\exp\left[-\frac{1}{\rho_0|V|}\right]\ ,\quad
\rho\equiv\frac{N}{\Omega}\ ,\quad
\rho_0 = \frac{mk_{\rm F}}{2\pi^2}
\label{eq53}
\end{equation}

From eqs.(\ref{eq38},\ref{eq53}) we conclude that the energy of the baryonic 
ground state is lower than for BCS state (for $n\ge4$ or, equivalently, 
$S\ge3/2$). Indeed, one usually assumes that a 
coupling constant is proportional to $1/n$ in the framework of large $n$ 
expansion, so we can put $V\equiv V_0/n$. Then the ratio of the baryonic and BCS 
binding energies can be expressed as follows:
$$
\frac{\Delta E_{\rm BCS}}{\Delta E_{\rm b}}
 = \exp\left[-\frac{1}{\rho_0|V|}+\frac{1}{\rho_0n|\tilde V|}\right]
 = \exp\left[-\frac{3n/8-1}{\rho_0|V_0|}\right] < 1\quad
 {\rm for}\quad n\ge4
$$
The last relation demonstrates the statement.

\section{Conclusion}

In this paper we have shown that, assuming that all scattering lengths
are approximately equal, the true ground state of a dilute gas of fermions with a 
high hyperfine spin is in fact of baryonic nature. The energy of the state and 
the gap in the spectrum (interpreting $\Delta$ by analogy with the BCS case) have 
been calculated. For comparison we calculated the 
energy of BCS $s$-wave ground states and showed that it is higher than
the energy of the baryonic ground state.

The possibility of forming a baryon-like bound state has been discussed in  
paper~\cite{Noz1}. The authors investigated the formation of a four-fermion 
($\alpha$-particle) condensate. While the results of \cite{Noz1} are very 
interesting they cannot be directly related to ours. Firstly, in the case of 
$\alpha$-particles $n=4$ and $1/n=1/4$ can be hardly regarded as a good 
expansion parameter to apply our results. Secondly, $n=4$ is the
dimension of the joint spin-isospin 
space and the Hamiltonian is not invariant under rotations in that space 
(exemplified by the absence of a di-neutron bound state, as against
the existence of the deuteron).

We note that if the interaction were strong then the quasi-baryons considered in 
the paper would become `real-space' composite bosons where the ground
state might be more appropriately described as being Bose-condensed. So 
it would interesting to study the evolution from weak to strong
coupling in the manner that Nozieres {\it et. al.} derived for an attractive 
fermion gas 
\cite{Noz}.

The method developed in this paper may be of 
interest in other fields such as the low-energy behavior of QCD 
\cite{Alkofer,NJL}, 
the quark-gluon plasma \cite{Harris} and neutron stars 
\cite{Prakash,Wilczek1,Wilczek2}.

\acknowledgments

This work was supported by EPSRC grant GR K68356.

\appendix

\section*{}

In this appendix we sketch the main steps of calculation of the norm of the 
baryonic ground state vector and averages (\ref{eq10},\ref{eq11}).

We consider the following generating functional
\begin{equation}
G(\bar\eta,\eta)\equiv
\langle\psi_{\rm b}|
\exp\left[\sum_{i=1}^n\sum_k c^\dagger_i(k)\eta_i(k)\right]
\exp\left[\sum_{i=1}^n\sum_k \bar\eta_i(k)c_i(k)\right]
|\psi_{\rm b}\rangle
\label{a1}
\end{equation}
from which we can obtain the norm and the required averages in an obvious way. 
Using the identity (\ref{eq5}) we can transform (\ref{a1}) to the form
\begin{eqnarray}
G(\bar\eta,\eta)\! &=&\! 
C \int {\rm D}\bar\xi{\rm D}\xi\ \exp\Biggl[
   \sum_{i=1}^{n-1}\sum_{k}\Bigl\{ - \bar\xi_i(k)\xi_i(k) + \bar\eta_i(k)\xi_i(k)
 + \bar\xi_i(k)\eta_i(k) - \bar\eta_i(k)\eta_i(k) \Bigr\}
\nonumber\\
&+&
   \sum_{k_1,\dots,k_n}\Bigl\{ 
   \bar\varphi_n(k_1,\dots,k_n)\eta_n(k_n) \xi_{n-1}(k_{n-1})\dots\xi_1(k_1)
\nonumber\\
&&
 + \varphi_n(k_1,\dots,k_n)\bar\xi_1(k_1)\dots\bar\xi_{n-1}(k_{n-1})    
   \bar\eta_n(k_n) \Bigr\}
\nonumber\\
&-&\!\!
   \sum_{k_1,\dots,k_n\atop p_1,\dots,p_n}
   \bar\varphi_n(k_1,\dots,k_n)\varphi_n(p_1,\dots,p_n) 
   \bar\xi_1(p_1)\xi_1(k_1)\dots \bar\xi_{n-1}(p_{n-1})\xi_{n-1}(k_{n-1})
   \delta_{k_n,p_n}
\Biggr]
\label{G}
\end{eqnarray}
Here we put $\alpha=n$ while it can be any number from $1$ to $n$. 

Consider first the norm of the ground state. Then after some transformations (see 
the main text) we obtain saddle point equations (\ref{sp1},\ref{sp2}). If we 
accept anzatz (\ref{anz}) then from (\ref{sp2}) we have
$$
s_3(k) = K_0\left|\frac{v(k)}{u(k)}\right|^2\ ,\qquad
s_4(k) = \frac{v(k)}{u(k)}
$$
\begin{eqnarray}
K_0\equiv K_0(k_1) &=& |B|^2\sum_{k_2,\dots,k_n}\delta_{k_1+\dots+k_n,0}
\prod_{i=2}^{n-1}\frac{|v(k_i)|^2s_1(k_i)}{|u(k_i)|^2}\cdot |v(k_n)|^2
\nonumber\\
c_4\equiv c_4(k_1,p_1)&=&
|B|^2\sum_{k_2,\dots,k_n\atop p_2,\dots,p_{n-1}}\delta_{k_1+\dots+k_n,0} 
\delta_{p_1+\dots+p_{n-1}+k_n,0}
\prod_{i=2}^{n-1}\frac{\bar v(k_i)v(p_i)}{\bar u(k_i)u(p_i)}\cdot |v(k_n)|^2
\nonumber\\
&&\times
\left[
   \prod_{i=2}^{n-1}\{\delta_{k_i,p_i}s_1(k_i)
   + c_2\bar s_2(p_i)s_2(k_i)\}
 - \prod_{i=2}^{n-1}\delta_{k_i,p_i}s_1(k_i)
\right]
\label{c4}
\end{eqnarray}
$K_0(k),c_4(k,p)$ are weakly dependent on their arguments and can be approximated 
by constants. On other hand, from (\ref{sp1},\ref{anz}) we obtain
$$
s_1(k) = \frac{1}{1+s_3(k)}\ ,\qquad
s_2(k) = \frac{s_4(k)}{1+s_3(k)}
$$
\begin{equation}
c_2 = - c_4\left[1 + c_4\sum_k\frac{|s_4(k)|^2}{1+s_3(k)}\right]^{-1}
\label{c2}
\end{equation}
Choosing $B$ such that $K_0=1$ we obtain the following solution
$$
s_1(k) = |u(k)|^2\ ,\qquad s_3(k) = \frac{|v(k)|^2}{|u(k)|^2}
$$
$$
s_2(k) = v(k)\bar u(k)\ ,\qquad s_4(k) = \frac{v(k)}{u(k)}
$$
It can be easily seen then that the condition $K_0=1$ is equivalent to the 
normalization condition (\ref{eq3}). From (\ref{c4},\ref{c2}) a relation for 
determining of the constant $c_4$ can be obtained. In the weak coupling 
approximation it takes the form:
$$
c_4 = \frac{{\rm const}}{n^3N_{\rm b}}
\left[ \frac{1}{(1+c_4N_{\rm b})^{n-2}} - 1 \right]\ ,\qquad
N_{\rm b} = \sum_k |v(k)|^2
$$
An obvious solution of this equation is $c_4=0$ and we obtain (\ref{sol}).

Consider now average (\ref{eq10}). Using the symmetry over spin we obtain from 
(\ref{G}) after integrating over the Grassmann fields:
\begin{eqnarray*}
&&\!\!\langle\psi_{\rm b}|c^\dagger_i(k)c_i(k)|\psi_{\rm b}\rangle =
\langle\psi_{\rm b}|c^\dagger_n(k)c_n(k)|\psi_{\rm b}\rangle =
 - \left.
   \frac{\partial}{\partial\eta_n(k)}\frac{\partial}{\partial\bar\eta_n(k)}
   G(\bar\eta,\eta)\right|_{\eta=\bar\eta=0}
\\
&&=
C \int{\rm D}\bar\sigma{\rm D}\sigma\
\sum_{k_1,\dots,k_{n-1}\atop p_1,\dots,p_{n-1}}
\bar\varphi_n(k_1,\dots,k_{n-1},k)\varphi_n(p_1,\dots,p_{n-1},k)
\prod_{i=1}^{n-1} [1+\bar\sigma_i]^{-1}(k_i,p_i)
\\
&&\qquad\times
\exp\Biggl[
 - \sum_{i=1}^{n-1}\sum_{k,p}\bar\sigma_i(k,p)\sigma_i(k,p)
 + \sum_{i=1}^{n-1}{\rm tr}\log[1+\bar\sigma_i]
 + K
\Biggr]
\end{eqnarray*}
A contribution of the first multiplier in the integrand to the saddle point 
equations is of the factorizable form and it can be shown that this contribution 
can be neglected. So we can use again solution (\ref{sol}). Hence
$$
\langle\psi_{\rm b}|c^\dagger_i(k)c_i(k)|\psi_{\rm b}\rangle = 
|v(k)|^2 |B|^2\sum_{k_1,\dots,k_{n-1}} \delta_{k_1+\dots+k_{n-1}+k,0}
\prod_{i=1}^{n-1}|v(k_i)|^2 = |v(k)|^2
$$
The last equality is due to the normalization condition (\ref{eq3}).

Finally we consider the average (\ref{eq11}). We have
\begin{eqnarray*}
&&\!\!\langle\psi_{\rm b}|\sum_q 
c^\dagger_i(k)c^\dagger_j(-k+q)c_j(-k'+q)c_i(k')
|\psi_{\rm b}\rangle =
\langle\psi_{\rm b}|
c^\dagger_1(k)c^\dagger_n(-k+q)c_n(-k'+q)c_1(k')
|\psi_{\rm b}\rangle
\\
&& =
\left.\sum_q
\frac{\partial}{\partial\eta_1(k)}\frac{\partial}{\partial\eta_n(-k+q)}
\frac{\partial}{\partial\bar\eta_n(-k'+q)}\frac{\partial}{\partial\bar\eta_1(k')}
G(\bar\eta,\eta)\right|_{\eta=\bar\eta=0}
 = \delta_{k,k'}|v(k)|^2\sum_q |v(q)|^2
\\
&& +
C \int{\rm D}\bar\sigma{\rm D}\sigma\
\sum_{k_1,\dots,k_{n-1},q\atop p_1,\dots,p_{n-1}}
\bar\varphi_n(k_1,\dots,k_{n-1},-k+q)\varphi_n(p_1,\dots,p_{n-1},-k'+q)
\prod_{i=1}^{n-1} [1+\bar\sigma_i]^{-1}(k_i,p_i) 
\\
&&\times
[1+\bar\sigma_1]^{-1}(k_1,k)[1+\bar\sigma_1]^{-1}(p_1,k')
\exp\Biggl[
 - \sum_{i=1}^{n-1}\sum_{k,p}\bar\sigma_i(k,p)\sigma_i(k,p)
 + \sum_{i=1}^{n-1}{\rm tr}\log[1+\bar\sigma_i]
 + K
\Biggr]
\end{eqnarray*}
Again it can be shown as before that we can use saddle solution (\ref{sol}), so 
we obtain
\begin{eqnarray*}
&&\!\!\!\langle\psi_{\rm b}|\sum_q 
c^\dagger_i(k)c^\dagger_j(-k+q)c_j(-k'+q)c_i(k')
|\psi_{\rm b}\rangle = \delta_{k,k'}|v(k)|^2\sum_q |v(q)|^2
\\
&& + \bar v(k)u(k) v(k')\bar u(k')
\sum_q \bar v(-k+q) v(-k'+q)
|B|^2\sum_{k_2,\dots,k_{n-1}} \delta_{k_2+\dots+k_{n-1}+q,0}
\prod_{i=2}^{n-1} |v(k_i)|^2 
\\
&&\approx
   \delta_{k,k'}|v(k)|^2\sum_q |v(q)|^2
 + \bar v(k)u(k) v(k')\bar u(k')\frac{1}{N_{\rm b}}\sum_q \bar v(-k+q) v(-k'+q)
\end{eqnarray*}
In the weak coupling limit $|v(k)u(k)|$ is non-zero only in a narrow region 
around the Fermi surface, so we can put $|k|=|k'|=k_{\rm F}$. Introducing the 
following notation
$$
f(\hat k,\hat k')\equiv
\left.\sum_q \bar v(-k+q) v(-k'+q)\right|_{|k|=|k'|=k_{\rm F}}
$$
we obtain (\ref{eq11}).


\begin{thebibliography}{99}

\bibitem{geist} W.~Geist, L.~You, T.~A.~B.~Kennedy, quant-ph/9810096.

\bibitem{optical} D.~M.~Stamper-Kurn, M.~R.~Andrews, A.~P.~Chikkatur, S.~Inouye, 
H.-J.~Miesner, J.~Stenger and W.~Ketterle, Phys.~Rev.~Lett., {\bf 80}, 2027 
(1998).

\bibitem{Cata}F.~S.~Cataliotti, E.~A.~Cornell, C.~Fort, M.~Inguscio, 
F.~Marin, M.~Prevedelli, L.~Ricci, G.~M.~Tino,  Phys. Rev. A {\bf 57},
1136 (1998).

\bibitem{Stoof} H.~T.~C.~Stoof, M.~Houbiers, C.~A.~Sackett,
R.~A.~Hulet, Phys. Rev. Lett. {\bf 76}, 10 (1996).

\bibitem{Hou1}M.~Houbiers, R.~Ferwerda, H.~T.~C.~Stoof,
W.~I.~McAlexander, 
C.~A.~Sackett, R.~A.~Hulet, Phys. Rev. A {\bf 56}, 4864 (1997) and 
M.~Houbiers, R.~Ferwerda, H.~T.~C.~Stoof,
W.~I.~McAlexander, 
C.~A.~Sackett, R.~A.~Hulet, Phys. Rev. A {\bf 57}, 4065 (1998).

\bibitem{bar}M.~A.~Baranov, D.~S.~Petrov, Phys. Rev. A, {\bf 58},
R801, (1998); M.~A.~Baranov, cond-mat/9801142.

\bibitem{Ho2} T.~L.~Ho and S.~K.~Yip, cond-mat/9808246.

\bibitem{Ho3} T.~L.~Ho and S.~K.~Yip, cond-mat/9810064.

\bibitem{nuc}{\it Nuclear Wallet Cards}, 5th edn., edited by Jagdish
K. Tuli, Brookhave National Laboratory.

\bibitem{Sutherland} B.~Sutherland, Phys. Rev. Lett. {\bf 20}, 98
(1968).

\bibitem{Tak} M.~Takahashi, Prog. Theor. Phys., {\bf 46}, 1388 (1971).

\bibitem{schlot1} P.~Schlottmann, J. Phys.: Condens. Matter, {\bf 5},
5869 (1993).

\bibitem{schlot2} P.~Schlottmann, J. Phys.: Condens. Matter, {\bf 6},
1359 (1994).

\bibitem{BCS} J.~Bardeen, L.~N.~Cooper and J.~R.~Schrieffer,
Phys. Rev. {\bf 108}, 1175 (1957).

\bibitem{Witten} E.~Witten, Nucl.~Phys., {\bf B160}, 57 (1979).

%\bibitem{BEC1} M.~H.~Anderson, J.~R.~Ensher, M.~R.~Mathews, C.~E.~Weiman and 
%E.~A.~Cornell, Science, {\bf 269}, 198 (1995).

%\bibitem{BEC2} C.~C.~Bradley, C.~A.~Sacket, J.~J.~Tollet,and R.~G.~Hulet, Phys. 
%Rev. Lett. {\bf 75}, 1687 (1995).

%\bibitem{BEC3}
%K.~B.~Davis, M.-O.~Mewes, M.~R.~Andrews, N.~J.~van~Druten, D.~S.~Durfee, 
%D.~M.~Kurn and W.~Ketterle, Phys.~Rev.~Lett., {\bf 75}, 3969 (1995).

%\bibitem{MIT} J.~Stenger, S.~Inouye, D.~M.~Stamper-Kurn, H.-J.~Miesner, 
%A.~P.~Chikkatur, and W.~Ketterle, Nature, {\bf 396}, 345 (1998).

\bibitem{AM} P.~W.~Anderson, P.~Morel, Phys.~Rev., {\bf 123}, 1911 (1961).

\bibitem{Ho1} T.~L.~Ho, Phys.~Rev.~Lett., {\bf 81}, 742 (1998).

\bibitem{Voll} D.~Vollhardt, P.~Wolfle, {\it The Superfluid Phases of Helium 3},
London: Taylor \& Francis (1990).

\bibitem{Gant} F.~R.~Gantmacher, {\it The Theory of Matrices}, v.2, New York: 
Chelsea Publishing Company (1959).

\bibitem{Noz1} G.~Ropke, A.~Schnell, P.~Schuck, P.~Nozieres, Phys. Rev. 
Lett. {\bf 80}, 3177 (1998).

\bibitem{Noz} P.~Nozieres, S.~Schmitt-Rink, J. Low Temp. Phys. {\bf 59}, 195 
(1985).

\bibitem{Alkofer} R.~Alkofer, H.~Reinhardt, H.~Weigel, Phys. Rep. {\bf 265}, 139 
(1996).

\bibitem{NJL} S.~P.Klevansky, Rev. Mod. Phys. {\bf 64}, 649 (1992).

\bibitem{Harris} J.~W.~Harris, B.~Muller, Ann. Rev. Nucl. Part. Sci. {\bf 46}, 
71 
(1996).

\bibitem{Prakash} M.~Prakash, I.~Bombachi, M.~Prakash, P.~J.~Ellis, 
J.~M.~Lattimer, R.~Knorren, Phys.Rep. {\bf 280}, 2 (1997).

\bibitem{Wilczek1} T.~Schafer, F.~Wilczek, {\it Continuity of Quark and Hadron 
Matter}, hep-ph/9811473.

\bibitem{Wilczek2} M.~Alford, K.~Rajagopal, F.~Wilczek, {\it Color-Flavor 
Locking 
and Chiral Symmetry Breaking in High Density QCD}, hep-ph/9804403.

\end{thebibliography}
\end{document}